\documentclass[preprint,aps,nofootinbib,11pt]{revtex4-1}
\usepackage{geometry}                
\usepackage{amsmath,amssymb,amsfonts,dcolumn,color,graphicx,graphics,latexsym,placeins,epsfig,tikz}
\usetikzlibrary{arrows,shapes}
\usepackage{subfigure,rotating,bm,mathrsfs}

\newcommand{\be}{\begin{equation}}
\newcommand{\ee}{\end{equation}}
\newcommand{\ba}{\begin{eqnarray}}
\newcommand{\ea}{\end{eqnarray}}

\usepackage[pagebackref=false, colorlinks=true]{hyperref}
\definecolor{redish}{rgb}{0.7,0.2,0.0}  
\definecolor{bluish}{rgb}{0.2,0.5,0.8}

\hypersetup{linkcolor=redish,          
                  citecolor=blue,        
                  filecolor=magenta,      
                  urlcolor=blue}          

\begin{document}
\author{Rajibul Shaikh}
\email{rshaikh@iitk.ac.in}
\affiliation{Department of Physics, \\ Indian Institute of Technology, \\ Kanpur 208016, India}
\title{\Large Black hole shadow in a general rotating spacetime obtained through Newman-Janis algorithm}

\begin{abstract}
The Newman-Janis (NJ) algorithm has been extensively used in the literature to generate rotating black hole solutions from nonrotating seed spacetimes. In this work, we show, using various constants of motion, that the null geodesic equations in an arbitrary stationary and axially symmetric rotating spacetime obtained through the NJ algorithm can be separated completely, provided that the algorithm is applied successfully without any inconsistency. Using the separated null geodesic equations, we then obtain an analytic general formula for obtaining the contour of a shadow cast by a compact object whose gravitational field is given by the arbitrary rotating spacetime under consideration. As special cases, we apply our general analytic formula to some known black holes and reproduce the corresponding results for black hole shadow. Finally, we consider a new example and study shadow using our analytic general formula.

\end{abstract}

\maketitle

\section{Introduction}
It is generally believed that the central supermassive compact region of our Galaxy and those of many other galaxies contain supermassive black holes. Images and shadows formed due to gravitational lensing of light provide an observational tool in probing the gravitational fields around such compact objects and in detecting their nature. The gravitational field near a black hole event horizon becomes so strong that its exterior geometry can possesses unstable circular photon orbits or unstable light rings (or a photon sphere in the case of a spherically symmetric, static black hole) which causes photons to undergo unboundedly large amount of bending (strong gravitational lensing) \citep{SL1,SL2,SL3,SL4,SL5}. A slight perturbation on photons on such unstable orbits can cause them to be either absorbed by the black hole or sent off to a faraway observer. Therefore, the event horizon of a black hole, together with the unstable light rings, is expected to create a characteristic shadowlike image (a darker region over a brighter background) of the photons emitted from nearby light sources or of the radiation emitted from an accretion flow around it. Very recently, the event horizon telescope (EHT) \cite{EHT1,EHT2,EHT3} has observed this shadow in the image of M87$^*$. However, the observational outcome of the image of the supermassive compact object Sagittarius A$^*$ (Sgr A$^*$) present at our Galactic center is yet to come.

While the intensity map of an
image depends on the details of the emission mechanisms of photons, the contour (silhouette) of the shadow is determined only by the spacetime metric itself, since it
corresponds to the apparent shape of the photon capture
orbits (or the unstable light rings) as seen by a distant observer. Therefore, strong lensing images and shadows offer us an exciting opportunity not only to detect the nature of a compact object but also to test whether or not the gravitational field around a compact object is described by the Schwarzschild or Kerr geometry. In light of this, there have been both analytic and numerical efforts to investigate shadows cast by different black holes in the last few decades. The shadow of a Schwarzschild black hole was studied by Synge \citep{synge_1966} and Luminet \citep{luminet_1979}. Bardeen studied the shadow cast by a Kerr black hole \citep{bardeen_1973} (see \cite{chandra} also). Consequently, the Kerr black hole shadow and its different aspects such as the measurement of the mass and spin parameter have been investigated by several authors \citep{falcke_2000,takahashi_2004,zakharov_2005b,beckwith_2005, broderick_2006a,takahashi_2007,hioki_2009,johannsen_2010, paolis_2011,kraniotis_2011,stuchlik_2018,kumar_2018,wei_2019}. The shadows cast by various other black holes have also been studied \citep{zakharov_2005a,zakharov_2014,stuchlik_2019,yumoto_2012, shipley_2016,gott_2019,young_1976,vries_2000,takahashi_2005, kraniotis_2014,moffat_2015,held_2019,cunha_2015,hioki_2008,dastan_2016, abdolrahimi_2015,li_2014,abdujabbarov_2016a,amir_2016,sharif_2016, saha_2018, wei_2013,abdujabbarov_2013,grenzebach_2014,amarilla_2012, eiroa,amarilla_2013,cunha_2017,atamurotov_2013b,atamurotov_2013, wang_2017,wang_2018,ovgun_2018,hou_2018,haroon_2019a,haroon_2019b, perlick_2018,mishra_2019,papnoi_2014,amir_2017,abdujabbarov_2015a, atamurotov_2015,abdujabbarov_2017,abdujabbarov_2015b,younsi_2016,held}. See \cite{shadow_review} for a recent brief review on shadows. Some recent studies, however, suggest that the presence of a shadow does not by itself prove that a compact object is necessarily a black hole. Other horizonless compact objects, which posses light rings around them, can also cast shadows \citep{shadow_horizon,broderick_2006b,bambi_2009,rajibul_2018a, bambi_2013a,ohgami_2015,mustafa_2015,ohgami_2016,rajibul_2018c, nedkova_2013,abdujabbarov_2016b,rajibul_2018b,gyulchev_2018,amir_2018, vincent_2016,abdikamalov_2019}.

Unlike the nonrotating ones, rotating black hole solutions are very hard to obtain as exact solutions of the field equations of various gravity theories. On the other hand, the Newman-Janis (NJ) algorithm provides an easier and more useful way to generate a stationary and axisymmetric rotating black hole spacetime from a static and spherically symmetric nonrotating seed metric \citep{NJ1,NJ2} (see \cite{NJ3} and \cite{NJ4} also). This method has been extensively used in the literature in recent times. In fact, many of the rotating black hole solutions cited above have been obtained through this method. The NJ method, however, has some shortcomings and may not be useful in some cases to generate the rotating black hole metric. Another useful method to derive general parametrization of axisymmetric black holes can be found in \cite{rezzolla_2014,rezzolla_2016}. In this work, we study shadow cast by a general rotating black hole generated from a nonrotating one through the NJ algorithm. A similar work has been considered in \cite{tsukamoto_2018} in the case when the initial nonrotating seed black hole spacetime has a particular ansatz. However, our work here is not restricted to such an ansatz and deals with a most general rotating black hole generated through the NJ algorithm.

This paper is organized as follows. In the next section, we briefly summarize the NJ algorithm and apply it to obtain a most general rotating black hole spacetime from a nonrotating one. In Sec. \ref{sec:shadow}, we separate null geodesic equations in the general rotating black hole spacetime and obtain a general analytic formula for obtaining the contour of the shadow which the black hole cast. We apply our formula to verify some known results in Sec. \ref{sec:example}. In Sec. \ref{sec:new_example}, we generate a new rotating black hole solution using the NJ algorithm and study its shadow. Finally, we conclude in Sec. \ref{sec:conclusions}.

\section{Rotating spacetime through Newman-Janis algorithm}
\label{sec:NJ}

In this section, we briefly summarize the NJ algorithm described in \cite{NJ1,NJ2} for the construction of a stationary and axisymmetric spacetime from a static and spherically symmetric one (see \cite{NJ3} also for more details). We start with the spherically symmetric, static spacetime given by
\begin{equation}
ds^2=-f(r) dt^2+\frac{dr^2}{g(r)}+h(r)\left(d\theta^2+\sin^2\theta d\phi^2\right).
\end{equation}
The first step of the algorithm is to write down the above metric in the advance null (Eddington-Finkelstein) coordinates $(u,r,\theta,\varphi)$ using the transformation
\begin{equation}
du=dt-\frac{dr}{\sqrt{fg}}.
\end{equation}
The metric in the advance null coordinates becomes
\begin{equation}
ds^2=-f(r) du^2-2\sqrt{\frac{f}{g}}dudr+h(r)\left(d\theta^2+\sin^2\theta d\phi^2\right).
\end{equation}
The second step is to express the inverse metric $g^{\mu\nu}$ using a null tetrad $Z_\alpha^\mu=(l^\mu,n^\mu,m^\mu,\bar{m}^\mu)$ in the form
\begin{equation}
g^{\mu\nu}=-l^\mu n^\nu -l^\nu n^\mu +m^\mu \bar{m}^\nu +m^\nu \bar{m}^\mu,
\end{equation}
where $\bar{m}^\mu$ is the complex conjugate of $m^\mu$, and the tetrad vectors satisfy the relations
\begin{equation}
l_\mu l^\mu = n_\mu n^\mu = m_\mu m^\mu = l_\mu m^\mu = n_\mu m^\mu =0,
\end{equation}
\begin{equation}
l_\mu n^\mu = - m_\mu \bar{m}^\mu =-1.
\end{equation}
One finds that the tetrad vectors satisfying the above relations are given by
\begin{equation}
l^\mu=\delta^\mu_r, \hspace{0.2cm} n^\mu=\sqrt{\frac{g}{f}}\delta^\mu_u-\frac{g}{2}\delta^\mu_r, \quad m^\mu=\frac{1}{\sqrt{2h}}\left(\delta^\mu_\theta+\frac{i}{\sin\theta}\delta^\mu_\phi\right).
\end{equation}
The third step is a complex transformation in the $r-u$ plane given by
\begin{equation}
r\rightarrow r'=r+ia\cos\theta, \hspace{0.3cm} u\rightarrow u'=u-ia\cos\theta,
\end{equation}
together with the complexification of the metric functions $f(r)$, $g(r)$ and $h(r)$. After the complex transformation, the new tetrad vectors become
\begin{equation}
l'^\mu=\delta^\mu_r, \quad n'^\mu=\sqrt{\frac{G(r,\theta)}{F(r,\theta)}}\delta^\mu_u-\frac{G(r,\theta)}{2}\delta^\mu_r,
\end{equation}
\begin{equation}
m'^\mu=\frac{1}{\sqrt{2H(r,\theta)}}\left(ia\sin\theta(\delta^\mu_u-\delta^\mu_r)+\delta^\mu_\theta+\frac{i}{\sin\theta}\delta^\mu_\phi\right),
\end{equation}
where $F(r,\theta)$, $G(r,\theta)$ and $H(r,\theta)$ are, respectively, the complexified form of $f(r)$, $g(r)$ and $h(r)$. Using the new tetrad, we find the new inverse metric using
\begin{equation}
g^{\mu\nu}=-l'^\mu n'^\nu -l'^\nu n'^\mu +m'^\mu \bar{m}'^\nu +m'^\nu \bar{m}'^\mu.
\end{equation}
The new metric in the advance null coordinates becomes
\begin{eqnarray}
ds^2&=&-Fdu^2-2\sqrt{\frac{F}{G}}dudr+2a\sin^2\theta\left(F-\sqrt{\frac{F}{G}}\right)du d\phi+2a\sqrt{\frac{F}{G}}\sin^2\theta drd\phi \nonumber \\
& &+H d\theta^2+\sin^2\theta\left[H+a^2\sin^2\theta\left(2\sqrt{\frac{F}{G}}-F\right)\right]d\phi^2.
\label{eq:null_coordinate_metric_1}
\end{eqnarray}
The final step of the algorithm is to write down the metric in Boyer-Lindquist form (where the only nonzero off diagonal term is $g_{t\phi}$) using the coordinate transformations
\begin{equation}
du=dt'+\chi_1(r)dr, \hspace{0.5cm} d\phi=d\phi'+\chi_2(r) dr.
\label{eq:transformation_to_BL}
\end{equation}
Inserting the above coordinate transformations in the metric (\ref{eq:null_coordinate_metric_1}) and setting $g_{t'r}$ and $g_{r\phi'}$ to zero, we obtain
\begin{equation}
\chi_1(r)=-\frac{\sqrt{\frac{G(r,\theta)}{F(r,\theta)}}H(r,\theta)+a^2\sin^2\theta}{G(r,\theta)H(r,\theta)+a^2\sin^2\theta},
\label{eq:chi1}
\end{equation}
\begin{equation}
\chi_2(r)=-\frac{a}{G(r,\theta)H(r,\theta)+a^2\sin^2\theta}.
\label{eq:chi2}
\end{equation}
Note that the transformation in (\ref{eq:transformation_to_BL}) is possible only when $\chi_1$ and $\chi_2$ depend only on $r$. If the right hand sides of Eqs. (\ref{eq:chi1}) and (\ref{eq:chi2}) depend on $\theta$ also, then we cannot perform a global coordinate transformation of the form (\ref{eq:transformation_to_BL}) \citep{NJ4}. Although it is not always possible to find a suitable complexification of the functions in such a way that $\chi_1$ and $\chi_2$ are independent of $\theta$, in many cases, it is. However, this involves a certain arbitrariness and an element of guess. There are many ways to complexify. Some are
\begin{equation}
\frac{1}{r}\rightarrow \frac{1}{2}\left(\frac{1}{r'}+\frac{1}{\bar{r}'}\right)=\frac{r}{\rho^2}, \quad r^2\rightarrow r'\bar{r}'=\rho^2,
\label{eq:complexify}
\end{equation}
where $\rho^2=r^2+a^2\cos^2\theta$. Finally, once the global coordinate transformation (\ref{eq:transformation_to_BL}) is allowed, the metric in the Boyer-Lindquist coordinate becomes
\begin{eqnarray}
ds^2 &=& -Fdt^2-2a\sin^2\theta\left(\sqrt{\frac{F}{G}}-F\right)dt d\phi+\frac{H}{GH+a^2 \sin^2\theta}dr^2+H d\theta^2 \nonumber \\
& & +\sin^2\theta\left[H+a^2\sin^2\theta\left(2\sqrt{\frac{F}{G}}-F\right)\right]d\phi^2,
\label{eq:final_general_metric}
\end{eqnarray}
where we have dropped the prime sign from $t'$ and $\phi'$. For later use, we define
\begin{equation}
\Delta(r)=G(r,\theta)H(r,\theta)+a^2\sin^2\theta,
\label{eq:Delta}
\end{equation}
\begin{equation}
X(r)=\sqrt{\frac{G(r,\theta)}{F(r,\theta)}}H(r,\theta)+a^2\sin^2\theta.
\label{eq:X}
\end{equation}
From Eqs. (\ref{eq:chi1}) and (\ref{eq:chi2}), note that $\Delta$ and $X$ must be independent of $\theta$ so that the transformation (\ref{eq:transformation_to_BL}) is allowed.

\section{Separation of null geodesic equations and black hole shadow}
\label{sec:shadow}

In this section, we separate the null geodesic equations in the general rotating spacetime (\ref{eq:final_general_metric}) using the Hamilton-Jacobi method and obtain a general formula for finding the contour of a shadow. The Hamilton-Jacobi equation is given by
\begin{equation}
\frac{\partial S}{\partial \lambda}+H=0, \quad H=\frac{1}{2}g_{\mu\nu}p^\mu p^\nu,
\label{eq:HJE}
\end{equation}
where $\lambda$ is the affine parameter, $S$ is the Jacobi action, $H$ is the Hamiltonian, and $p^\mu$ is the momentum defined by
\begin{equation}
p_{\mu}=\frac{\partial S}{\partial x^\mu}=g_{\mu\nu}\frac{dx^\nu}{d\lambda}.
\label{eq:momentum}
\end{equation}
Since the metric tensor $g_{\mu\nu}$ and hence the Hamiltonian $H$ is independent of the coordinates $t$ and $\phi$, we have two constants of motion. These are the conserved energy $E=-p_t$ and the conserved angular momentum $L=p_\phi$ (about the axis of symmetry). If there is a separable solution of Eq. (\ref{eq:HJE}), then, in terms of the already known constants of the motion, it must take the form
\begin{equation}
S=\frac{1}{2}\mu ^2 \lambda - E t + L \phi + S_{r}(r)+S_{\theta}(\theta),
\label{eq:action_ansatz}
\end{equation}
where $\mu$ is the mass of the test particle. For a photon, we take $\mu=0$. Putting Eq. (\ref{eq:action_ansatz}) in the Hamilton-Jacobi equation, we obtain after some simplifications
\begin{eqnarray}
-\left(GH+a^2\sin^2\theta\right)& &\left(\frac{dS_r}{dr}\right)^2+\frac{\left[\left(\sqrt{\frac{G}{F}}H+a^2\sin^2\theta\right)E-aL \right]^2}{\left(GH+a^2\sin^2\theta\right)}-(L-aE)^2 \nonumber \\
& & =\left(\frac{dS_\theta}{d\theta}\right)^2+L^2\cot^2\theta-a^2E^2\cos^2\theta.
\label{eq:S1S2}
\end{eqnarray}
Note that, since the quantities $\left(GH+a^2\sin^2\theta\right)[=\Delta(r)]$ and $\left(\sqrt{\frac{G}{F}}H+a^2\sin^2\theta\right)[=X(r)]$ are functions of $r$ only [see Eqs. (\ref{eq:Delta}) and (\ref{eq:X})], the left- and right-hand side of Eq. (\ref{eq:S1S2}) are only functions of $r$ and $\theta$, respectively. Therefore, each side must be equal to a separation constant. After separation, we obtain
\begin{equation}
-\left(GH+a^2\sin^2\theta\right)\left(\frac{dS_r}{dr}\right)^2+\frac{\left[\left(\sqrt{\frac{G}{F}}H+a^2\sin^2\theta\right)E-aL \right]^2}{\left(GH+a^2\sin^2\theta\right)}-(L-aE)^2= \mathcal{K},
\end{equation}
\begin{equation}
\left(\frac{dS_\theta}{d\theta}\right)^2+L^2\cot^2\theta-a^2E^2\cos^2\theta=\mathcal{K},
\end{equation}
where the separation constant $\mathcal{K}$ is known as the Carter constant. Using Eq. (\ref{eq:momentum}), we obtain the following separated geodesic equations for the photon:
\begin{equation}
\frac{F}{G}\Delta(r)\frac{dt}{d\lambda}=\left[H+a^2\sin^2\theta\left(2\sqrt{\frac{F}{G}}-F\right)\right]E-a\left(\sqrt{\frac{F}{G}}-F\right)L,
\label{eq:t_eqn}
\end{equation}
\begin{equation}
\frac{F}{G}\Delta(r)\sin^2\theta\frac{d\phi}{d\lambda}=a\sin^2\theta\left(\sqrt{\frac{F}{G}}-F\right)E+FL,
\label{eq:phi_eqn}
\end{equation}
\begin{equation}
H\frac{dr}{d\lambda}=\pm \sqrt{R(r)},
\label{eq:r_eqn}
\end{equation}
\begin{equation}
H\frac{d\theta}{d\lambda}=\pm \sqrt{\Theta(\theta)},
\label{eq:theta_eqn}
\end{equation}
where
\begin{equation}
R(r)=\left[X(r)E-aL\right]^2-\Delta(r)\left[\mathcal{K}+\left(L-aE\right)^2\right],
\end{equation}
\begin{equation}
\Theta(\theta)=\mathcal{K}+a^2E^2\cos^2\theta-L^2\cot^2\theta,
\end{equation}
and $\Delta(r)$ and $X(r)$ are defined, respectively, in Eqs. (\ref{eq:Delta}) and (\ref{eq:X}). Note that $R(r)$ and $\Theta(\theta)$ must be non-negative; i.e., we must have
\begin{equation}
\frac{R(r)}{E^2}=\left[X(r)-a\xi\right]^2-\Delta(r)\left[\eta+\left(\xi-a\right)^2\right]\geq 0,
\label{eq:R}
\end{equation}
\begin{equation}
\frac{\Theta(\theta)}{E^2}=\eta+(\xi-a)^2-\left(\frac{\xi}{\sin\theta}-a\sin\theta\right)^2\geq 0
\label{eq:Theta}
\end{equation}
for the photon motion, where $\xi=L/E$ and $\eta=\mathcal{K}/E^2$.

The unstable circular photon orbits in the general rotating spacetime must satisfy $R(r_{ph})=0$, $R'(r_{ph})=0$ and $R''\geq 0$, where $r=r_{ph}$ is the radius of the unstable photon orbit. The first two conditions give
\begin{equation}
\left[X(r_{ph})-a\xi\right]^2-\Delta(r_{ph})\left[\eta+\left(\xi-a\right)^2\right]=0,
\label{eq:Req0}
\end{equation}
\begin{equation}
2X'(r_{ph})\left[X(r_{ph})-a\xi\right]-\Delta'(r_{ph})\left[\eta+\left(\xi-a\right)^2\right]=0.
\label{eq:Rpeq0}
\end{equation}
After eliminating $\eta$ from the last two equations and solving for $\xi$, we obtain
\begin{equation}
\xi=\frac{X(r_{ph})}{a} \quad \text{or} \quad \xi=\frac{X(r_{ph})\Delta'(r_{ph})-2\Delta(r_{ph})X'(r_{ph})}{a\Delta'(r_{ph})}.
\label{eq:xisol}
\end{equation}
Out of these two solutions for $\xi$, only one is valid for the purpose of describing a black hole shadow. If we take the first solution $\xi=X/a$, then from Eq. (\ref{eq:Req0}), we find that the corresponding solution for $\eta$ is given by
\begin{equation}
\eta+(\xi-a)^2=0,
\end{equation}
which is compatible with the requirement $\Theta(\theta)\geq 0$ [see Eq. (\ref{eq:Theta})] only for
\begin{equation}
\theta=\theta_{ph}=\text{a constant}, \quad \text{and} \quad \xi=a\sin^2\theta_{ph},
\end{equation}
when $\Theta(\theta)=0$ for $\theta=\theta_{ph}$. This case is similar to the one of the cases of the Kerr black hole \citep{chandra}. This set of solutions for $\xi$ and $\eta$ represents principal null-geodesics and can not describe a black hole shadow. Therefore, to describe a black hole shadow, we consider the second solution of $\xi$ given in Eq. (\ref{eq:xisol}). Using this second solution, we solve for $\eta$ from Eq. (\ref{eq:Rpeq0}). We obtain
\begin{equation}
\xi=\frac{X_{ph}\Delta'_{ph}-2\Delta_{ph}X'_{ph}}{a\Delta'_{ph}},
\label{eq:xi}
\end{equation}
\begin{equation}
\eta=\frac{4a^2X'^2_{ph}\Delta_{ph}-\left[\left(X_{ph}-a^2\right)\Delta'_{ph}-2X'_{ph}\Delta_{ph} \right]^2}{a^2\Delta'^2_{ph}}.
\label{eq:eta}
\end{equation}
where the subscript ``$ph$" indicates that the quantities are evaluated at $r=r_{ph}$. Equations (\ref{eq:xi}) and (\ref{eq:eta}) give the general expressions for the critical impact parameters $\xi$ and $\eta$ of the unstable photon orbits which describe the contour of a shadow.

A particular case of the above study with
\begin{equation}
f(r)=g(r)=1-\frac{2m(r)}{r}, \quad h(r)=r^2
\label{eq:particular}
\end{equation}
has been considered in \cite{tsukamoto_2018}. In this case, using (\ref{eq:complexify}), the metric functions can be complexified to obtain \citep{NJ4}
\begin{equation}
F=G=1-\frac{2m(r)r}{\rho^2}, \quad H=\rho^2, \quad \rho^2=r^2+a^2\cos\theta^2.
\label{eq:special1}
\end{equation}
Therefore, $\Delta(r)$ and $X(r)$ become
\begin{equation}
\Delta(r)=r^2-2m(r)r+a^2,\quad X(r)=r^2+a^2,
\label{eq:special2}
\end{equation}
which are functions of $r$ only. Using Eqs. (\ref{eq:special1}) and (\ref{eq:special2}), it is straightforward to show that the geodesic equations (\ref{eq:t_eqn})--(\ref{eq:theta_eqn}) as well as the expressions for $\xi$ and $\eta$ given in Eqs. (\ref{eq:xi}) and (\ref{eq:eta}) exactly match with those obtained in \cite{tsukamoto_2018}. However, in our most general case here, we do not restrict the metric functions to be of the form given in Eq. (\ref{eq:particular}).

The unstable photon orbits form the boundary of a shadow. The apparent shape of a shadow are obtained by using the celestial coordinates $\alpha$ and $\beta$ which lie in the celestial plane perpendicular to the line joining the observer and the center of the spacetime geometry. The coordinates $\alpha$ and $\beta$ are defined by \citep{celestial}
\begin{equation}
\alpha=\lim_{r_0\to\infty}\left(-r_0^2\sin\theta_0\frac{d\phi}{dr}\Big\vert_{(r_0,\theta_0)}\right),
\end{equation}
\begin{equation}
\beta=\lim_{r_0\to\infty}\left(r_0^2\frac{d\theta}{dr}\Big\vert_{(r_0,\theta_0)}\right),
\end{equation}
where $(r_0,\theta_0)$ are the position coordinates of the observer. We consider that the general metric is asymptotically flat. Therefore, $F\sim 1$, $G\sim 1$, $H\sim r^2$, $\Delta\sim r^2$ and $X\sim r^2$ in the limit $r\to \infty$. After taking the limit, we obtain
\begin{equation}
\alpha=-\frac{\xi}{\sin\theta_0},
\label{eq:alpha}
\end{equation}
\begin{equation}
\beta=\pm \sqrt{\eta+a^2\cos^2\theta_0-\xi^2\cot^2\theta_0}.
\label{eq:beta}
\end{equation}
The shadows are constructed by using the unstable photon orbit radius $r_{ph}$ as a parameter and then plotting parametric plots of $\alpha$ and $\beta$ using Eqs. (\ref{eq:xi}), (\ref{eq:eta}), (\ref{eq:alpha}) and (\ref{eq:beta}).

\section{Some known examples}
\label{sec:example}
\subsection{Kerr, Kerr-Newman, and tidally charged rotating braneworld black hole}

The Kerr \citep{kerr,NJ1}, the Kerr-Newman \citep{NJ2} and the tidally charged rotating braneworld black hole \citep{tidal} are, respectively, rotating solutions of Einstein-vacuum equations, Einstein-Maxwell equations and the effective field equations of the Randall-Sundrum braneworld in vacuum. Through the NJ algorithm, these black hole solutions can be obtained from the nonrotating metric given by
\begin{equation}
f(r)=g(r)=1-\frac{2M}{r}+\frac{q}{r^2},\quad h(r)=r^2,
\end{equation}
where $q=0$ represents Schwarzschild black hole, $q=Q^2$ represents electrically charged Reissner-Nordstrom black hole with $Q$ being the electric charge, and $q=-Q^2_{*}$ represents tidally charged braneworld black hole with $Q_{*}$ being the tidal charge. Using (\ref{eq:complexify}), the metric functions in this case can be complexified as \citep{NJ1,NJ2}
\begin{equation}
F=G=1-\frac{2Mr}{\rho^2}+\frac{q}{\rho^2}, \quad H=\rho^2, \quad \rho^2=r^2+a^2\cos^2\theta.
\end{equation}
Using these complexified functions in Eqs. (\ref{eq:Delta}) and (\ref{eq:X}), we obtain
\begin{equation}
\Delta(r)=r^2-2Mr+a^2+q, \quad X(r)=r^2+a^2.
\end{equation}
The black hole horizons are given by $\Delta=0$. When $(M^2-a^2-q)\geq 0$, we have Kerr black hole for $q=0$, Kerr-Newman black hole for $q=Q^2$ and tidally charged braneworld black hole for $q=-Q^2_{*}$. The shadows cast by these black holes have already been studied \citep{bardeen_1973,young_1976,vries_2000,takahashi_2005,amarilla_2012}. Using the above expression for $\Delta(r)$ and $X(r)$ in Eqs. (\ref{eq:xi}) and (\ref{eq:eta}), we obtain
\begin{equation}
\xi=\frac{2r_{ph}(2Mr_{ph}-q)-(r_{ph}+M)(r_{ph}^2+a^2)}{a(r_{ph}-M)},
\end{equation}
\begin{equation}
\eta=\frac{4a^2r_{ph}^2(Mr_{ph}-q)-r_{ph}^2\left[r_{ph}(r_{ph}-3M)+2q\right]^2}{a^2(r_{ph}-M)^2},
\end{equation}
which are the same as those of the Kerr ($q=0$), Kerr-Newman ($q=Q^2$) and tidally charged braneworld black hole ($q=-Q^2_{*}$) obtained, respectively, in \cite{bardeen_1973}, \cite{vries_2000} and \cite{amarilla_2012}. Note that the authors in \cite{amarilla_2012} have taken $q=Q$ with $Q<0$ for the tidally charged braneworld black hole. 

\subsection{Kerr-Sen black hole}
The Kerr-Sen black hole is a rotating charged black hole solution obtained in heterotic string theory \citep{sen}. In \cite{yazadjiev_2000}, the author has shown that, using the NJ algorithm, the Kerr-Sen black hole solution can be obtained from the spherically symmetric, static metric given by
\begin{equation}
f(r)=g(r)=\frac{1-\frac{r_1}{r}}{1+\frac{r_2}{r}},\quad h(r)=r^2\left(1+\frac{r_2}{r}\right),
\end{equation}
where $r_1$ and $r_2$ are related to the mass $M$ and the electric charge $Q$ by $r_1+r_2=2M$ and $r_2=Q^2/M$. Note however that $M$ and $Q$ in \cite{sen} are related to the two parameters $m$ and $\alpha$ by $M=(m/2)(1+\cosh\alpha)$ and $Q=(m/\sqrt{2})\sinh\alpha$. In this case, using (\ref{eq:complexify}), the metric functions can be complexified as \citep{yazadjiev_2000}
\begin{equation}
F=G=\frac{1-\frac{r_1r}{\rho^2}}{1+\frac{r_2r}{\rho^2}},\quad H=\rho^2\left(1+\frac{r_2r}{\rho^2}\right), \quad \rho^2=r^2+a^2\cos^2\theta.
\end{equation}
Therefore, $\Delta(r)$ and $X(r)$ in this case become
\begin{equation}
\Delta(r)=r^2-r_1r+a^2, \quad X(r)=r^2+r_2r+a^2.
\end{equation}
The shadows for this black hole have been studied in \cite{hioki_2008}. Using the last equation in Eqs. (\ref{eq:xi}) and (\ref{eq:eta}), we obtain
\begin{equation}
\xi=\frac{(r_1+r_2)(r_{ph}^2-a^2)-(2r_{ph}+r_2)(r_{ph}^2-r_1r_{ph}+a^2)}{a(2r_{ph}-r_1)},
\end{equation}
\begin{eqnarray}
\eta &=& \frac{r_{ph}^2}{a^2(2r_{ph}-r_1)^2}\{4a^2(r_1+r_2)(2r_{ph}+r_2) \nonumber\\
& & -\left[(2r_{ph}+r_2)(r_{ph}-r_1)-(r_1+r_2)r_{ph}\right]^2\}.
\end{eqnarray}
We find that the above expressions for the critical impact parameters $\xi$ and $\eta$ are the same as those obtained in Eq. (27) of \cite{hioki_2008}, after we replace $r_2=r_0$ and $r_1=2M-r_0$ in the above expressions for $\xi$ and $\eta$.

\section{A new example: rotating dilaton black hole and its shadow}
\label{sec:new_example}

\begin{figure}[h]
\centering
\subfigure[$~a/M=0.5$, $Q_M/M=0.4$]{\includegraphics[scale=0.6]{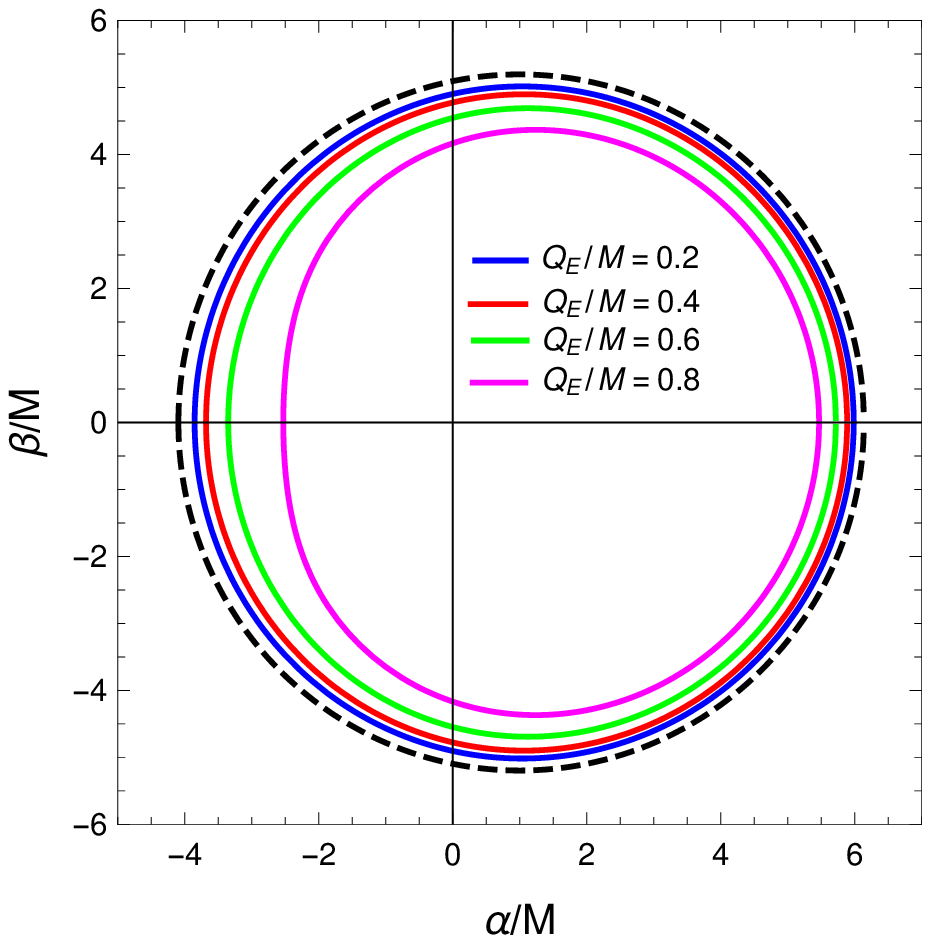}}
\subfigure[$~a/M=0.7$, $Q_M/M=0.4$]{\includegraphics[scale=0.6]{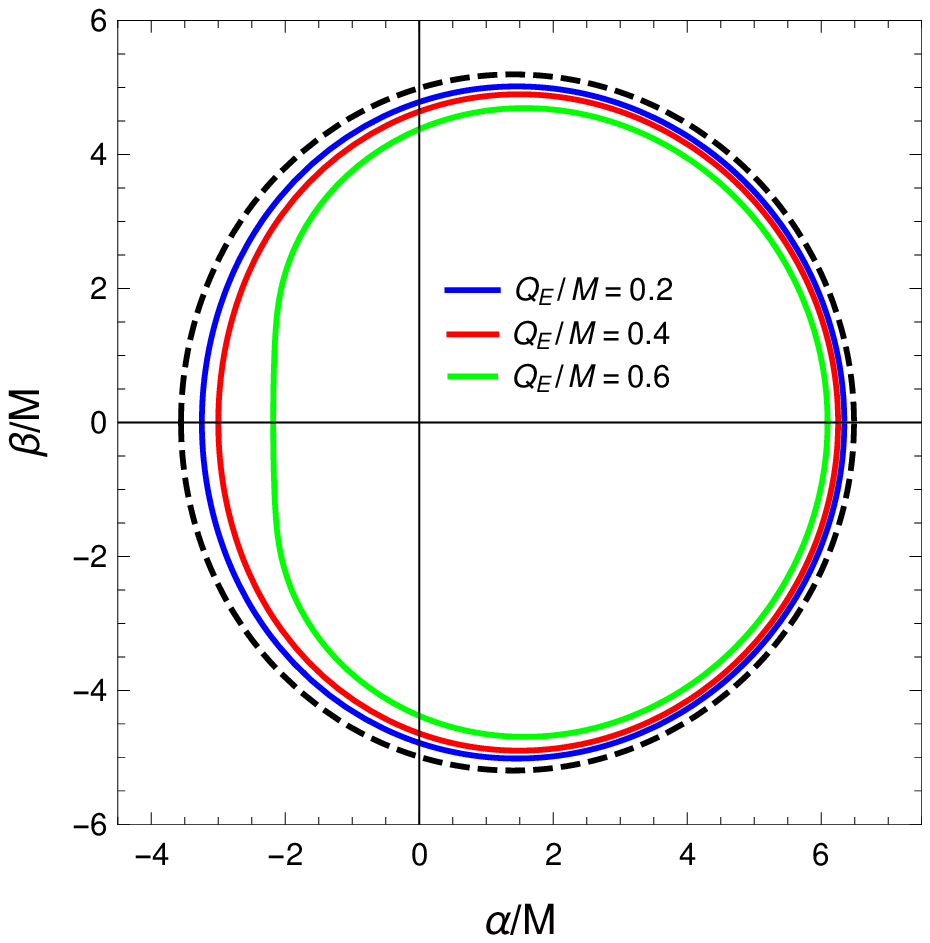}}
\subfigure[$~Q_E/M=0.3$, $Q_M/M=0.2$]{\includegraphics[scale=0.65]{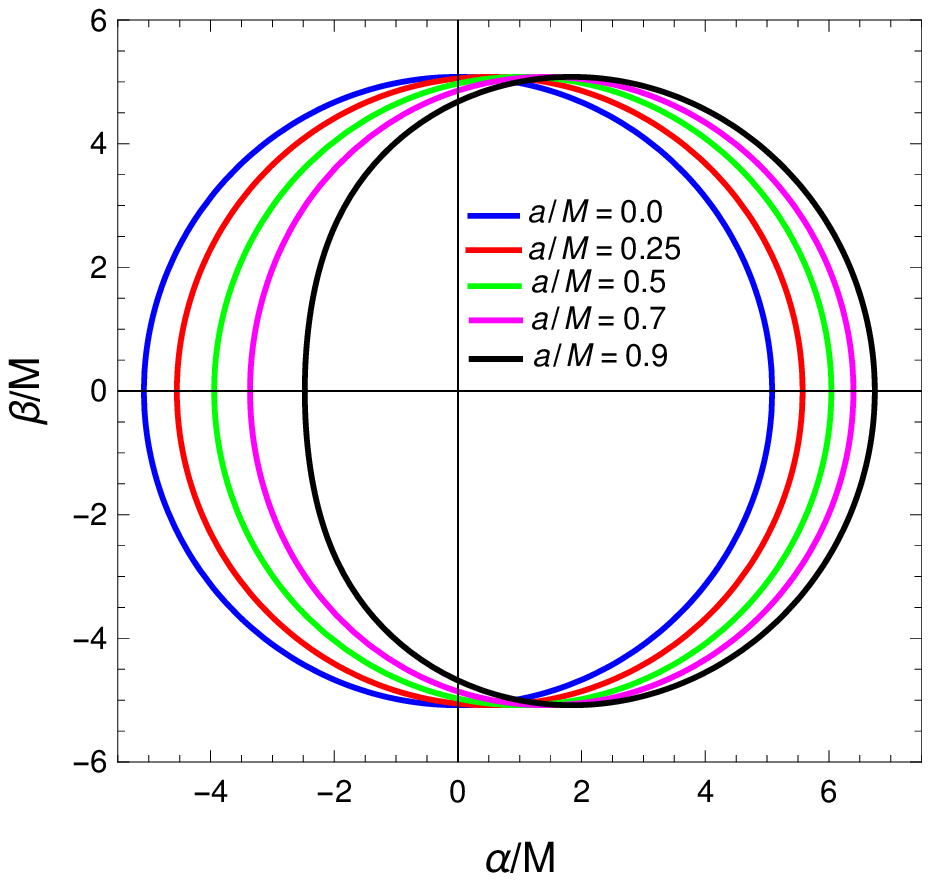}}
\caption{Shadows cast by a rotating charged dilaton black hole for different parameter values [(a)-(c)]. The dashed contours are for a Kerr black hole with the spin values shown in the corresponding plot. The inclination angle of the observer is $\theta_0=\pi/2$.} 
\label{fig:NJA_shadow}
\end{figure}

We now consider a new example to bring out the usefulness of our general analytic formula for obtaining a shadow. We consider the nonrotating charged dilaton black hole given by \citep{new1,new2}
\begin{equation}
f(r)=g(r)=\frac{\left(1-\frac{r_{-}}{r}\right)\left(1-\frac{r_{+}}{r}\right)}{1-\frac{r_0^2}{r^2}},\quad h(r)=r^2\left(1-\frac{r_0^2}{r^2}\right),
\end{equation}
\begin{equation}
r_\pm=M\pm \sqrt{M+r_0^2-Q_E^2-Q_M^2}, \quad r_0=\frac{Q_M^2-Q_E^2}{2M},
\end{equation}
and apply the NJ algorithm to obtain the corresponding rotating charged dilaton black hole. Here, $M$ is the mass, and $Q_E$, $Q_M$ and $r_0$ are, respectively, related to the electric, the magnetic and the dilaton charge. To complexify the function, we first write $f(r)$ and $g(r)$ as
\begin{equation}
f(r)=g(r)=\frac{\left(1-\frac{2M}{r}+\frac{q}{r^2}\right)}{1-\frac{r_0^2}{r^2}},
\end{equation}
where we have replaced $(r_{-}+r_{+})=2M$ and have defined $q=r_{-}r_{+}=(Q_E^2+Q_M^2-r_0^2)$. Using (\ref{eq:complexify}), we now complexify the functions in the following way:
\begin{equation}
F=G=\frac{\left(1-\frac{2Mr}{\rho^2}+\frac{q}{\rho^2}\right)}{1-\frac{r_0^2}{\rho^2}},\quad H=\rho^2\left(1-\frac{r_0^2}{\rho^2}\right),\quad \rho^2=r^2+a^2\cos^2\theta.
\end{equation}
Using above equations, we find that the expressions for $\Delta(r)$ and $X(r)$ become
\begin{equation}
\Delta(r)=r^2-2Mr+a^2+q, \quad X(r)=r^2-r_0^2+a^2.
\end{equation}
Therefore, the critical impact parameters of the null geodesics in this case become
\begin{equation}
\xi=\frac{2r_{ph}(2Mr_{ph}-q)-r_0^2(r_{ph}-M)-(r_{ph}+M)(r_{ph}^2+a^2)}{a(r_{ph}-M)},
\end{equation}
\begin{eqnarray}
\eta &=& \frac{1}{a^2(r_{ph}-M)^2}\{4a^2r_{ph}^2(Mr_{ph}-q)-4a^2r_0^2r_{ph}(r_{ph}-M) \nonumber\\
& & -\left[r_{ph}^2(r_{ph}-3M)+2qr_{ph}+r_0^2(r_{ph}-M)\right]^2\}.
\end{eqnarray}
Note that when $Q_E=Q_M$, i.e., $r_0=0$, the dilaton charge vanishes, and the above results match with those of the Kerr-Newman black hole obtained in the previous section. To the best of our knowledge, construction of the above rotating metric using the NJ algorithm and its shadows have not been considered before. Figure \ref{fig:NJA_shadow} shows the shadows cast by the above rotating dilaton black hole for different values of the parameters.

\section{Conclusions}
\label{sec:conclusions}
After the very recent observation of the very first image of the black hole M87$^*$ \citep{EHT1,EHT2,EHT3}, a black hole shadow will continue to be an important probe of spacetime structure and gravity in the strong curvature regime. In this work, we have studied the shadow cast by an arbitrary stationary, axially symmetric and asymptotically flat rotating black hole spacetime generated through the NJ algorithm. To this end, we have completely separated the null geodesic equations using different constants of motion and obtained an analytic general formula which can be used to find the contour of the shadow cast by a rotating black hole. To demonstrate the usefulness of our general analytic formula, we have applied it to some known examples and reproduced the corresponding results. Finally, we have considered a new example and studied its shadows. Our analytic formula will be useful to obtain more new results for black hole shadows. Though we have applied our analytic formula to study shadows cast by rotating black holes, it can also be used to study shadows cast by other compact objects.

\end{document}